# Open GOP Resolution Switching in HTTP Adaptive Streaming with VVC

Robert Skupin, Christian Bartnik, Adam Wieckowski, Yago Sanchez, Benjamin Bross, Cornelius Hellge, Thomas Schierl

Video Communication and Applications Department
Fraunhofer Institute for Telecommunications—Heinrich Hertz Institute
{firstname.lastname}@hhi.fraunhofer.de

*Abstract*—The user experience in adaptive HTTP streaming relies on offering bitrate ladders with suitable operation points for all users and typically involves multiple resolutions. While open GOP coding structures are generally known to provide substantial coding efficiency benefit, their use in HTTP streaming has been precluded through lacking support of reference picture resampling (RPR) in AVC and HEVC. The newly emerging Versatile Video Coding (VVC) standard supports RPR, but only conversational scenarios were primarily investigated during the design of VVC. This paper aims at enabling usage of RPR in HTTP streaming scenarios through analysing the drift potential of VVC coding tools and presenting a constrained encoding method that avoids severe drift artefacts in resolution switching with open GOP coding in VVC. In typical live streaming configurations, the presented method achieves -8.57% BD-rate reduction compared to closed GOP coding while in a typical Video on Demand configuration, -1.89% BD-rate reduction is reported. The constraints penalty compared to regular open GOP coding is 0.65% BD-rate in the worst case. The presented method was integrated into the publicly available open source VVC encoder VVenC v0.3.

## I. Introduction

HTTP streaming has become an important path of video distribution over the last decade and over-the-top service providers can reach hundreds of millions of users via the public internet today. Standard protocols such as Dynamic Adaptive Streaming over HTTP (DASH) [1] enable a service provider to stream media to clients by having a server offering the media at various bitrates in a temporally segmented form. A client device is then able to download successive segments for continuous playback by selecting amongst the offered variants of a particular segment according to the available network bandwidth and its decoding capabilities in a dynamic and adaptive fashion. In practice, the content is offered as multiple so-called representations generated by optimized bitrate ladders, which often involve multiple resolutions and fidelities in order to optimize the perceived quality for a particular bitrate and thereby user experience [2]. Since each segment is typically coded without dependencies to earlier segments using so-called closed Group-Of-Pictures (GOP) coding structures [3], the downloaded and depacketized segment data can be concatenated to a conforming bitstream and fed into a decoder. Opposed to such closed GOP structures, segments using so-called open GOP coding structures contain some pictures that employ inter-prediction from pictures in earlier segments which benefits coding efficiency. While the pictures using inter-prediction from earlier segments can be skipped from being output without playback issues or visual artefacts when random accessing a segment as they come first in presentation order, an issue arises when a resolution switch occurs during continuous playout as skipping these pictures leads to a non-seamless switch. Even when switching between representations of equal resolution using open GOP structures, some pictures may be dropped or exhibit severe visual artefacts when segments are not encoded properly for switching.

Proliferated earlier generation codecs such as AVC [4] and HEVC [5] do not offer reference picture resampling (RPR) functionality required to use reference pictures of different resolution. Therefore, after resolution switching, when performed at open GOP structures, some pictures of a segment cannot be correctly decoded as reference pictures from earlier segments are not available in the required resolution, which results in non-constant frame rate playout at the segment switch from dropped pictures. In [6], the authors presented approaches to overcome the issue of open GOP resolution switching by either employing normative changes to the HEVC decoding process or using the less proliferated scalable extension of HEVC (SHVC) that offers RPR functionality. However, the available solutions have not led to substantial adoption of open GOP coding in HTTP streaming up to now.

The recently finalized version 1 of the Versatile Video Coding (VVC) standard [7] is the latest video coding standard that emerged from the collaborative work of the Video Coding Expert Group of ITU-T and the Sub Committee 29 of ISO/IEC also known as Moving Picture Experts Group. Aside offering substantially increased coding efficiency compared to earlier generation codecs [8], VVC also includes many application-driven features in the initial Main 10 profile such as RPR. During VVC development, RPR was mainly investigated in the context of conversational scenarios with low-delay coding structures [9] where real-world requirements on latency and buffer sizes set tight limits for the feasibility of insertion of intra coded pictures for resolution switching.

However, RPR in VVC can also provide substantial benefit to coding efficiency in video encoding for the streaming domain. This paper presents a method of constrained encoding to enable open GOP resolution switching in HTTP streaming using VVC. Experiments investigating the general coding efficiency impact of open GOP coding structures in VVC as well as the picture quality impact at segment switches are reported. The remainder of the paper is organized as follows. Section II provides an overview of GOP coding structures and segmentation. Section III describes the proposed constrained encoding method. Section IV details the experiments and reports corresponding results which is followed by a conclusion in Section V.

## II. Coding Structures and Segmentation

This section provides an overview of structures within a VVC bitstream and media segmentation for streaming. Media segments are generally aligned with intra random access point (IRAP) pictures using intra coding tools only. IRAP pictures may appear frequently in a coded video bitstream to allow functionalities such as seeking or fast forwarding, but also serve as switching points for adaptive HTTP streaming. Systems for Video on Demand (VoD) streaming typically align segments with IRAP picture periods, i.e., IRAP pictures are typically placed at the segment start and the desired segment duration determines the temporal distance between IRAP pictures. However, there are use-cases, e.g., very low delay streaming, in which not all media segments contain an IRAP picture, so that small segments can be made available for transmission without needing to wait for an IRAP picture and thus reduce the latency at the content generation side. Segment sizes may vary in length depending on the target application. For instance, VoD services allow players to build larger buffers (e.g., 30 seconds) to overcome throughput fluctuations for which segment sizes up to several seconds can be reasonable design choice [3]. However, live services that require more stringent end-to-end delays do not allow such large buffers at the client side and hence require more frequent switching points and shorter segments of 1 second or less.

Pictures between two IRAP pictures are typically encoded in a bi-predicted hierarchical GOP structure involving reordering before presentation as far as decoding delay requirements allow because such a structure provides substantial coding efficiency benefit as introduced in AVC [10]. The hierarchy structure of a GOP can be used for temporal scalability in which decoding all pictures up to a given layer correspond to a given framerate and a corresponding Temporal Id (Tid) value is assigned to each picture as shown in Fig. 1 for a GOP size of 8 pictures. A GOP can be defined as all pictures from a first Tid 0 picture up to but not including the following Tid 0 picture in decoding order. Typically, segments include one or more GOP structures depending on IRAP period and GOP size. While in HEVC, the amount of reference picture slots in the Decoded Picture Buffer (DBP) allowed GOP sizes of up to 16 pictures, DPB capacity was increased in VVC allowing hierarchical GOP sizes of up to 32 pictures.

Pictures following an IRAP picture in decoding order but preceding it in presentation order were introduced in HEVC and referred to as leading pictures. They can be further distinguished into Random Access Skipped Leading (RASL) and Random Access Decodable Leading (RADL). While RADL pictures may only use reference pictures from the corresponding IRAP picture onwards in decoding order, RASL pictures may use reference pictures preceding the IRAP in addition. IRAP pictures of the Instantaneous Random Access (IDR) type reset the DBP and can only have leading pictures of the RADL picture type leading to so-called closed GOP structures. Further IRAP pictures of the Clean Random Access (CRA) type, on the other hand, do not reset the DPB. Hence, reconstructed pictures from before the CRA picture in decoding order are available as reference for future pictures, i.e. the RASL pictures allowing for so-called open GOP coding structures. RASL pictures exhibit an increased coding efficiency compared to RADL pictures but can be rendered undecodable when reference pictures are not available, e.g. during a random access at the associated IRAP at the beginning of a segment without decoding the previous segment. A more detailed overview of the high-level syntax of VVC can be found in [11].

When the reference pictures of a RASL picture are located in the previous segment and the streaming client switches representations after such a previous segment, the client decoder will decode the RASL picture using a different variant of at least part of the reference pictures compared to encoder side. Such a situation can lead to a non-conforming bitstream if the content is not generated appropriately or to significant mismatches in the reconstructed RASL pictures and this drift may propagate to all RASL pictures up to but not including the associated CRA picture. The following section discusses appropriate generation of content to allow using open GOP structures while maintaining bitstream conformance at segment switches and avoiding undesirable drift that would be detrimental to visual quality during switches.

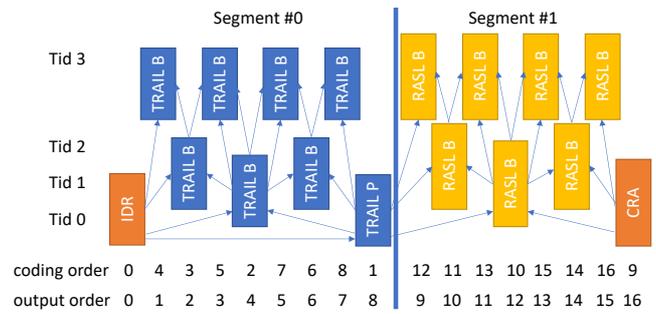

Fig. 1. Two successive segments with different resolution in which the second segment employs an open GOP coding structure with reference pictures from the first segment.

### III. CONSTRAINED OPEN GOP ENCODING

The numerous inter-prediction tools in VVC exhibit varying potential to cause drift when open GOP switching is carried out and at the same time, tool usage is bound by conformance constraints. This section analyses the drift potential of the inter-prediction tools in VVC at open GOP resolution switching and proposes a constrained encoding method to overcome severe artefacts of open GOP resolution switching while ensuring VVC conformance.

*A. Drift Potential of VVC Coding Tools*

A first set of coding tools in VVC can be categorized as sample-to-sample prediction, e.g. regular block-based translatory motion-compensated sample prediction known from many predecessors of VVC or a newly introduced inter-prediction mode in VVC referred to as affine motion compensation (AMC) which decomposes a prediction block into smaller sub-blocks that are individually motion-compensated to emulate affine motion compensation [12]. Prediction Refinement with Optical Flow (PROF) as an optional component of AMC or Bi-directional Optical Flow (BDOF) are further newly introduced inter-prediction tools in VVC that alter predicted samples by relying on optical flow-based methods in order to emulate sample-wise inter-prediction. When a different representation is used as reference for reconstruction using such sample-to-sample prediction tools, the visual quality of the reconstructed pictures will lean towards the visual quality of said representation and away from the visual quality of the original representation. However, such sample-to-sample prediction has a comparatively low potential to cause visually disturbing artefacts but rather leads to a graceful quality transition in a given sequence of RASL pictures as prediction source samples of a first visual quality are progressively updated through residual information at a second visual quality.

A second set of coding tools in VVC is used for syntax (i.e. model parameter) prediction, either from syntax or samples of a picture. Similar to its predecessors, VVC allows for motion vector (MV) prediction on block basis using temporal MV candidates from a so-called collocated reference picture via Temporal Motion Vector Prediction (TMVP) [13]. This feature was extended in VVC by introducing a finer granular TMVP variant on sub-block basis (SBTMVP) adding a displacement step in finding the corresponding motion information in the collocated reference picture. Further tools in this second set can be characterized as sample-to-syntax prediction tools. A completely new inter-prediction tool introduced tool in VVC is Decoder-side Motion Vector Refinement (DMVR) which refines the accuracy of MVs in bi-prediction based on the mirroring property of two reference pictures with equal and opposing temporal distance to the current picture. A further new tool in VVC is Cross Component Linear Model (CCLM) that allows to intra-predict the chroma components of a block from the respective luma component using a linear model wherein model parameters are derived from the reconstructed luma sample values [14]. Specifically, model parameters for chroma prediction are derived from extrema of the neighbouring luma and chroma sample values which can result in severe drift from even a single sample with outlier value. An even further new tool was introduced to the loop filtering stage of VVC and is referred to Luma Mapping and Chroma Scaling (LMCS) in which chroma sample values undergo a scaling using parameters derived from the luma samples. However, chroma scaling parameters are derived from the average value of a larger number of luma samples with mitigates drift potential compared to CCLM. Errors in predicted MVs from syntax-to-syntax and sample-to-syntax inter-prediction tools have comparatively high potential to lead to severe artefacts in the subsequent sample-to-sample prediction tools that use these erroneously predicted MVs as spatial or temporal MV candidates. The inherent error propagation over subsequent pictures may lead to increasing artefact magnitude. This is especially valid for (SB)TMVP which exhibits the most visible artefacts in open GOP switching but similarly applies to DMVR. Same is also valid for other prediction models such as CCLM, that are carried out based on parameters derived from the reconstructed sample values. Figure 2 illustrates the effect of general syntax or parameter prediction errors on the visual and objective quality of the luma and chroma sample values in a RASL picture using a GOP size of 32 pictures in regular open GOP coding even without resolution change at the switching point.

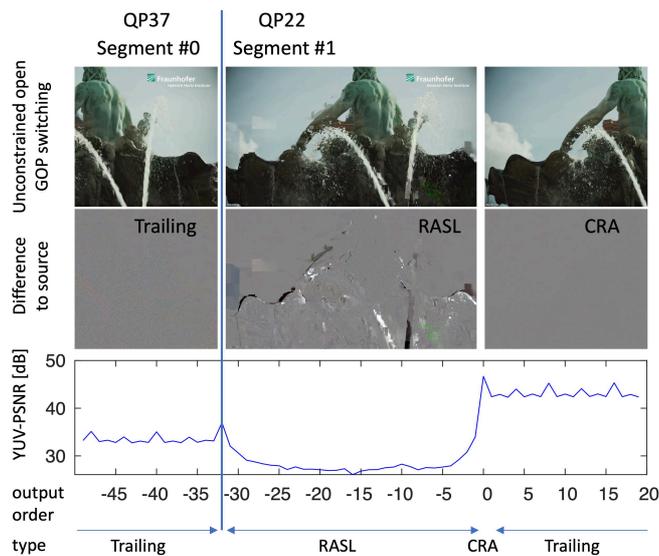

Fig. 2. Effect of regular open GOP segment switching on RASL picture quality using test sequence NeptuneFountain3 and a GOP size of 32 pictures.

A third issue in open GOP switching can arise from usage of Adaptation Parameter Sets (APSs) in VVC which carry filter coefficients for the Adaptive Loop Filter (ALF), parameters for Luma Mapping with Chroma Scaling (LMCS) and quantization scaling lists. RASL pictures may refer to APSs transmitted before the respective CRA in decoding order which are available during continuous decoding but are unavailable when random accessing at the CRA picture which is unproblematic since the associated RASL pictures are dropped in this case. However, open GOP resolution switching can cause references to unavailable APSs of earlier segment variants which crashes non-error-resilient decoders or creates visual artefacts when using parameters of wrong APSs with coincidentally matching identifiers. Similar to syntax prediction tools, this issue has a high potential to create a visual disturbance up to complete decoder failures.

B.   *Proposed Constrained Encoding Method*

In order to prevent the issues described above when performing open GOP resolution switching, this paper proposes a constrained encoding method for VVC that consists of three pillars.

First, the RASL pictures associated with a CRA are constrained so that no picture preceding the CRA in decoding order is selected as collocated reference picture to perform syntax-to-syntax prediction such as (SB)TMVP. Thereby, decoder side uses the exact same reference pictures for motion information prediction as on encoder side and any syntax prediction error from incorrect source motion information through earlier segment pictures is prevented. In our implementation, the first RASL picture in decoding order is limited to use only its associated CRA picture as collocated reference picture which naturally hosts only zero motion vectors while further RASL pictures also have access to non-zero temporal MV candidates of preceding RASL pictures. With respect to sample-to-syntax prediction tools, DMVR is disabled for all RASL pictures. Thereby, erroneous sample values of reference pictures from earlier segments that differ from encoder-side or drift-affected samples of other RASL pictures do not cause errors in the sample-to-syntax prediction chain. Albeit having lower drift potential, further tools have to be constrained for all RASL pictures with reference pictures preceding the associated CRA in coding order to ensure VVC conformance after a segment switch with resolution change. Usage of the optical flow related tools BDOF and PROF are disabled as well as motion-compensated prediction wrapping around the vertical picture boundaries which is a tool targeted at 360-degree video coding. In addition, a further new feature of VVC, namely independently coded subpictures, which are useful in viewport dependent streaming of 360-degree video, has to be disabled to use RPR. All of the above tool constraints are also part of conformance constraints defined in the VVC specification for pictures to use RPR (e.g. output order 9, 10 and 11 in Fig. 1) and are applied also to RASL pictures without RPR in the proposed encoding method to limit drift propagation. Beyond the conformance constraints associated with RPR usage, a further tool constraint is necessary as prediction techniques which employ parameter prediction from reconstructed samples may cause noticeable artefacts. Out of this category, CCLM is disabled through encoder-side constraints for RASL pictures in our implementation, as VVC version 1 syntax only allows sequence-wise disabling that comes at significant coding performance impact.

Second, the APSs for all pictures in coding order onwards from the CRA picture need to be present within the respective segment. Therefore, all parameters for processing of ALF, LMCS and quantization scaling lists are conveyed within the CRA picture at the beginning of each segment. In our implementation, the processing related to ALF, LMCS and quantization scaling lists is reset in a similar fashion as for closed GOP coding structures.

Third, from the perspective of VVC high level syntax, the individual encoding of variants has to be carried out in a coordinated fashion with the target of open GOP switching on decoder side in mind. Therefore, the Sequence Parameter Sets (SPSs) of all segment variants need to be aligned so that segment switching does not trigger the start of a new coded layer video sequence through changes in the SPS. For instance, with proper coordination, the SPS should indicate the maximum resolution within the bitstream ladder, matching block sizes and chroma formats, a proper matching level indicator and the relevant constraint flags such as gci_no_res_change_in_clvs_constraint_flag, sps_ref_pic_resampling_enabled_flag and sps_res_change_in_clvs_allowed_flag with appropriate configuration to enable usage of RPR on decoder side. Devices with lower capabilities than required for the indicated maximum resolution or level need to be served with properly adjusted SPSs through system layer mechanisms.

RPR in VVC has been designed in a constrained manner to limit its implementation and runtime complexity as evident from the above tool constraint discussion. An important aspect in this complexity consideration is that memory bandwidth while accessing scaled reference samples in RPR usage is acceptable and not significantly higher than without RPR. Coded pictures in VVC are accompanied with a so-called scaling window that is used to determine the scaling factor between two pictures. In order to set a bound to the memory bandwidth requirements of RPR, relation of scaling windows of pictures using RPR and scaling windows of their reference pictures is limited to allow an eightfold upscaling and a twofold downscaling at maximum. In other words, assuming each scaling window matches the picture size of its representation, it is allowed to use RPR when switching to a representation that has eightfold higher pictures sizes. However, down-switching may only use RPR if the picture size decreases by no more than half in each dimension. In adaptive streaming scenarios, up-switching is typically carried out in a progressive manner, i.e. increasing resolution or quality gradually over segments. However, when it comes to down-switching, it might happen that when the buffer of a player is starving, the player switches to the lowest quality to avoid buffer underruns, which means that downward switches likely do not happen progressively. One way to mitigate this limitation of RPR in VVC is to encode the lowest quality representation with closed GOP structures so that it can serve as a fallback for when picture sizes decrease to less than half during such non-progressive down-switching events.

## IV. EXPERIMENTS AND RESULTS

The experiments in this paper are carried out with an optimized open-source implementation of the open-source VVC encoder VVenC v0.3.1 [15] that is publicly available under BSD-type license and offers substantial runtime enhancement compared to the official reference software VTM [16]. Seven test sequences with a length of 10 seconds and 60 fps from the publicly available 8K Berlin test sequences suite were used in the experiments [17]. The general encoding configuration is aligned with the common test conditions as used in VVC development [18], i.e. encoding with constant quantization parameters (QPs) of 22, 27, 32 and 37 in a random access configuration. Three resolutions (2160p, 1080p and 720p), two GOP sizes (32 and 16 pictures) and three IRAP periods or segment lengths (64, 128 and 256 pictures) were investigated.

### A. Constrained Open GOP Coding Performance

Table I reports per-component and joint overall BD-rate gains of general open GOP coding structures and the presented constrained open GOP method, both relative to employing closed GOP coding structures for different GOP sizes, IRAP periods and resolutions. It can be seen that the average coding gain scales roughly linearly with GOP size and IRAP period as both directly determine the amount of affected RASL pictures in the bitstream. It is further noticeable that open GOP coding gains decrease with resolution. This effect presumably originates from the increasing amount of residual information with increased spatial resolution. For operation points such as GOP size 32 and IRAP period 64 which are suitable for low delay streaming with short segment sizes, substantial gains of roughly -9.22% BD-rate can be achieved with regular open GOP coding while a longer IRAP period 256 as suitable for VoD still achieves -2.35% BD-rate gain. The results for GOP size 16 and IRAP period 64 (-5.17% BD-rate gain) roughly match the results reported for open GOP coding with HEVC and SHVC in [6]. The presented constrained open GOP coding method in this paper is able to retain a major fraction of the regular open GOP coding gains with a negligible penalty of roughly 0.65% BR-rate for the best performing configuration and less for the other configurations, ultimately achieving up to -8.57% BD-rate reduction compared to closed GOP structures.

TABLE I. OPEN GOP CODING AND PRESENTED CONSTRAINED OPEN GOP CODING VS. CLOSED GOP CODING

| Res. | Temp. Config. | | Results vs closed GOP | | | | | |
|---|---|---|---|---|---|---|---|---|
| | | | Open GOP | | | Constr. open GOP | | |
| | GOP | IRAP | Y | U | V | Y | U | V |
| 2160p | 32 | 64 | -8.71% | -10.7% | -10.3% | -8.07% | -9.53% | -8.83% |
| 1080p | | | -9.12% | -11.1% | -10.7% | -8.60% | -9.57% | -8.73% |
| 720p | | | -9.38% | -11.3% | -10.8% | -8.87% | -9.93% | -9.08% |
| Overall | | | -9.07% | -11.1% | -10.6% | -8.51% | -9.68% | -8.88% |
| | | | -9.22%[a] | | | -8.57%[a] | | |
| 2160p | 32 | 128 | -4.26% | -4.92% | -4.64% | -3.82% | -4.22% | -3.88% |
| 1080p | | | -4.52% | -4.91% | -4.63% | -4.11% | -4.00% | -3.58% |
| 720p | | | -4.68% | -4.90% | -4.75% | -4.24% | -4.00% | -3.76% |
| Overall | | | -4.49% | -4.91% | -4.67% | -4.06% | -4.07% | -3.74% |
| | | | -4.49%[a] | | | -4.02%[a] | | |
| 2160p | 32 | 256 | -2.23% | -2.43% | -2.26% | -1.91% | -1.96% | -1.74% |
| 1080p | | | -2.40% | -2.32% | -2.14% | -2.05% | -1.67% | -1.41% |
| 720p | | | -2.49% | -2.31% | -2.30% | -2.10% | -1.70% | -1.72% |
| Overall | | | -2.37% | -2.35% | -2.23% | -2.02% | -1.78% | -1.62% |
| | | | -2.35%[a] | | | -1.98%[a] | | |
| 2160p | 16 | 64 | -5.24% | -5.44% | -5.10% | -4.81% | -4.85% | -4.33% |
| 1080p | | | -5.29% | -4.84% | -4.57% | -4.90% | -4.03% | -3.48% |
| 720p | | | -5.27% | -4.76% | -4.69% | -4.86% | -3.97% | -3.59% |
| Overall | | | -5.26% | -5.01% | -4.79% | -4.86% | -4.28% | -3.80% |
| | | | -5.17%[a] | | | -4.72%[a] | | |
| 2160p | 16 | 128 | -2.49% | -2.40% | -2.20% | -2.17% | -1.98% | -1.74% |
| 1080p | | | -2.55% | -2.04% | -1.87% | -2.22% | -1.47% | -1.10% |
| 720p | | | -2.55% | -1.88% | -1.85% | -2.18% | -1.36% | -1.25% |
| Overall | | | -2.53% | -2.11% | -1.98% | -2.19% | -1.60% | -1.36% |
| | | | -2.45%[a] | | | -2.09%[a] | | |
| 2160p | 16 | 256 | -1.29% | -1.12% | -1.03% | -1.02% | -0.85% | -0.65% |
| 1080p | | | -1.33% | -0.94% | -0.80% | -1.03% | -0.46% | -0.33% |
| 720p | | | -1.33% | -0.82% | -0.79% | -0.97% | -0.39% | -0.41% |
| Overall | | | -1.32% | -0.96% | -0.87% | -1.01% | -0.56% | -0.46% |
| | | | -1.26%[a] | | | -0.94%[a] | | |

[a] averaged per-resolution BD-rates based on YUV-PSNR using weights of 6 (Y), 1 (U) and 1 (V).

*B. RASL Picture Quality*

A further experiment investigates the quality of drift affected RASL pictures after segment switching with the proposed constrained encoding method. For this purpose, up- and down-switching between two adjacent resolutions is simulated at all possible switching points and upscaled YUV-PSNR in the higher resolution is measured at the switch point. Figure 3 shows the results for GOP size 32 and IRAP period 64 averaged over all switching points and QPs as well as upper and lower bound YUV-PSNR values of the respective open and closed GOP variants without switching. It can be seen that in the up-switching case, the quality exhibits a linear trend in transitioning between the two quality levels with an average YUV-PSNR of RASL pictures of 1.77dB below the quality of the high-resolution open GOP variant and 2.82dB above the low-resolution closed GOP variant. However, in the down-switching case, the results indicate a abrupt quality degradation compared to up-switching behaviour with an average YUV-PSNR of 3.72dB below the high-resolution open GOP variant and 0.87dB above the low-resolution closed GOP variant. The sharp quality degradation in down-switching occurs at the first RASL picture in display order which exhibits a drop in YUV-PSNR of 2.92dB with respect to the corresponding picture of the open GOP high-resolution variant. It is asserted that the asymmetry between up- and down-switching behaviour originates from the quality loss associated to downscaling of high-quality reference pictures and should be less pronounced in quality switching within a single resolution.

## V. CONCLUSION

This paper investigates open GOP resolution switching in HTTP streaming using VVC and presents a constrained encoding method to avoid severe artefacts in resolution switching. Experiments investigated the overall coding gain benefit of -8.57% BD-rate reduction for the proposed constrained encoding method compared to traditional closed GOP coding structures in live streaming scenarios and a BD-rate reduction of -1.98% for VoD scenarios. It is worth noting that the increased DPB size of VVC compared to HEVC and the resulting larger GOP sizes allow for roughly doubling the open GOP related coding gains seen with HEVC. Additionally, the quality transition of RASL pictures in the YUV-PSNR domain during up- and down-switching events was investigated. It was shown that while quality increases almost linearly in up-switching, down-switching results in a relatively abrupt quality degradation which is asserted to originate from downscaling of high-quality reference pictures. It is also worth noting that the RPR related conformance constraints related to BDOF and PROF may be omitted when targeting QP switching only as they do not induce severe artefacts but only need to be applied to enable RPR usage. The presented constrained encoding method for open GOP switching was integrated into VVenC v0.3.

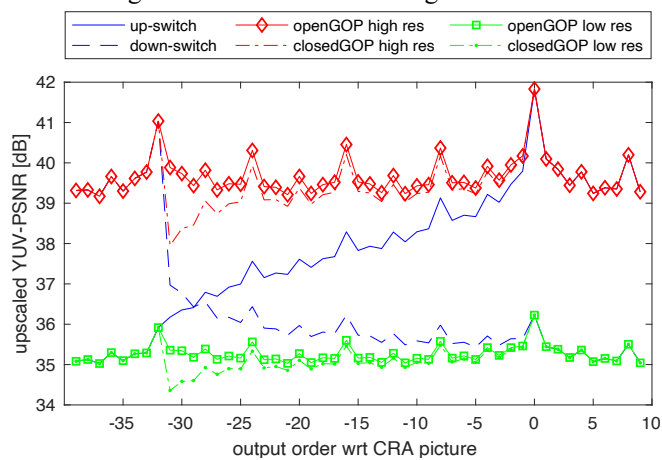

Fig. 3. Quality transition of the presented constrained encoding method for openGOP resolution switching with GOP size 32 and averaged over sequences, switching points and QPs.